\documentclass[twocolumn,showpacs,preprintnumbers,amsmath,amssymb]{revtex4}

\usepackage{graphicx}
\usepackage{dcolumn}
\usepackage{bm}
    \usepackage{amssymb}

\begin{document}



\title{Interface Trap Density Metrology of state-of-the-art undoped Si n-FinFETs}%
\author{Giuseppe Carlo Tettamanzi*, Abhijeet Paul*, Sunhee Lee, Saumitra R. Mehrotra, Nadine Collaert, Serge Biesemans, Gerhard Klimeck, Sven Rogge}


\begin{abstract}

The presence of interface states at the MOS interface is a well-known cause of device degradation. This is particularly true for ultra-scaled FinFET geometries where the presence of a few traps can strongly influence device behavior. Typical methods for interface trap density ($D_{it}$) measurements are not performed on ultimate devices, but on custom designed structures. We present the first set of methods that allow direct estimation of $D_{it}$ in state-of-the-art FinFETs,  addressing a critical industry need.

\end{abstract}
\maketitle 




\section{Introduction} \label{sec:I}

\begin{figure}[!t]
\centering
\includegraphics[width=3.5 in,height=1.85in]{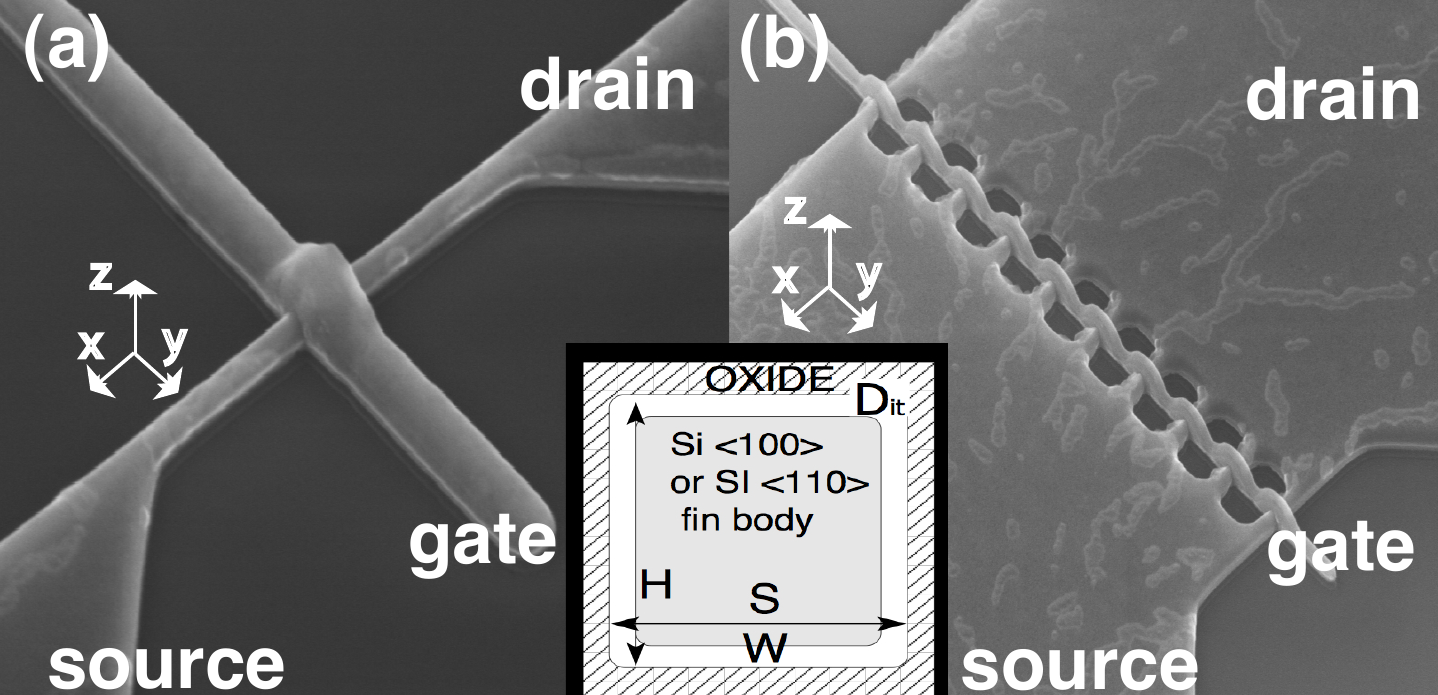}
\caption{Scanning-electron-microscope (SEM) image of a state-of-the-art FinFET device with the channel having a $<$100$>$ orientation and a single fin. b) SEM image of a FinFET device with the channel having a $<$110$>$ orientation and ten fins. In the inset the schematic of the cross-sectional cut in the Y-Z plane of a typical trigated FinFET is shown. The active cross-section ($S$) is in grey, $H$, $W$, and $D_{it}$ are the physical height, the physical width and the interface trap density, respectively.}
\label{fig:Finfet_cartoon}
\end{figure}

Non-planar trigated FinFET geometry (Fig.~\ref{fig:Finfet_cartoon}a, b) provides a viable solution to the channel length ($L_{ch}$) scaling due to their better gate to channel electrostatic coupling and reduced Short Channel Effects (SCEs) \cite{Won133}. In a recent work \cite{Tet150}, it has been demonstrated that by using thermionic emission, it is possible to measure (1) the active channel cross-section area ($S$) (inset Fig.~\ref{fig:Finfet_cartoon}), which represents the portion of the physical cross-section of channel where the charge flows, and (2) the source to channel barrier height ($E_{b}$), which reflects on the ease with which electrons travel from the source (drain) to the channel, hence opening new ways to investigate FinFETs. Furthermore, it was found that, although the trends of $S$ in the experimental and the simulated data were identical, differences in the absolute values were observed. These differences were found to be caused by the presence of interface states at the metal-oxide-semiconductor interface of the experimental devices \cite{Tet150, Lee186}. These states can trap electrons and enhance the presence of screening, therefore reducing the action of the gate on the channel, and as a final result, a decrease in the absolute value of $S$ in the experimental data is observed. Here we show that, by using simple mathematical manipulations and the difference between experimental and simulated values of $S$ and of the capacitive coupling $\alpha=\frac{dE_{b}}{dV_{g}}$ \cite {Tet150},  it is possible to infer the interface trap density ($D_{it}$). Typical $D_{it}$ measurements are not performed on ultimate devices but on custom designed structures \cite{Kap232}. Such custom structures may only be partially reflective for the possibly surface orientation-dependent and geometry-dependent $D_{it}$. Here, we provide a simple set of methods for the direct estimation of $D_{it}$ in ultimate devices. The comparison between the values of $D_{it}$ obtained with our two methods and between our results and the results obtained using a method implemented in the past \cite{Kap232} show similar trend. A new approach to trap density metrology is of critical importance as CMOS scaling leads device dimensions into nanometer regime. At these scales, quantities such as $D_{it}$ can vary rapidly with device geometry, rendering old techniques inadequate as they cannot be applied directly in these ultra-scaled devices.

\begin{table}[!t]

\begin{center}
\resizebox{8.9cm}{1.5cm}{
\begin{tabular}{|c|c|c|c|c|c|c|}
\hline
 & & &  & && \\
 \textbf{FinFET label} & \textbf{W(nm)} &\textbf{H(nm)}  & \textbf{L(nm)} &\textbf{Number of channels}&\textbf{Channel orientation} & \textbf{$H_2$ anneal }\\
 & & & &&&  \\
\hline
A & 25  &65 & 40&$1$&$<100>$& Yes \\
\hline
 B& 25&65&40&$1$&$ <100>$& No\\
 \hline  
 
 C &5&65&40&$1$&$ <100>$&No\\
 \hline

D and E &18&40&40&$10$&$ <110>$&Yes\\
 \hline

F   &$\sim$ 3-5&40&40&$10$&$<110>$&Yes\\
 \hline    
\end{tabular}}
\end{center}
 \caption{Details of the n-FinFETs used in this study and the labels used for them. All FinFETs have undoped channel and n-doped source and drain. }
\label{table1}
\end{table}


\section{Device and experimental details}\label{Sec:III}

The undoped n-FinFETs used in this work ($A-F$, see table \ref{table1}) consist of nanowire channels etched on a Si intrinsic film with a wrap-around gate covering three faces of the channels (Fig. \ref{fig:Finfet_cartoon}) \cite{Col108}. An HfSiO layer isolates a TiN layer from the intrinsic Si channel \cite{Col108}. Devices with the same channel length, ($L$ = 40 nm), different channel height ($H$ = 40 nm and 65 nm), different channel widths (3 nm $\lesssim$ $W$ $\lesssim$ 25 nm), $<$100$>$ or $<$110$>$ channel orientation and different surface treatment ($A, D, E, F$ with hydrogen anneal step during fabrication \cite{Lee186} and $B, C$ without) have been studied. Differential conductance ($G=\partial I_{SD}/\partial V_{SD}$) data are taken at $V_{SD}$ = $0$ $V$ using a lock-in technique. The complete experimental procedure to extract $S$ and $E_b$ can be found in Ref.~\cite{Tet150}.

\section{Modeling Approach}\label{sec:II}


To obtain the self-consistent charge and potential in n-FinFETs, the electronic structure is calculated using an atomistic 10 band $sp^3d^5s^*$ semi-empirical Tight-Binding (TB) \cite{Kli601,Boy115201}, which captures the geometrical and potential confinement, takes into account the atomic positions in the device \cite{Kli601,Boy115201,Neo1286}, and is coupled self-consistently to a $2D$ Poisson solver \cite{Neo1286,Paul1}. The thermionic current in the FinFETs is obtained using a ballistic top-of-the-barrier (ToB) model \cite{Neo1286,Paul1}. Due to the extensively large cross-section of the devices that combines up to 44,192 atoms (for $H$ = 65 nm, $W$ = 25 nm FinFETs) in the simulation domain, a new NEMO 3D code has been integrated in the top of the barrier analysis \cite{Sunhee1}. Using thermionic fitting procedure \cite{Tet150}, $E_{b}$, $\alpha$ and $S$ can be extracted using the experimental and theoretical conductance ($G$) in the thermionic emission regime for a 3D system \cite{sze1} as,  

\begin{equation}
\label{Gbulk}
   G_{3D} = SA^{*}T\frac{e}{k_{B}}exp\Big(-\frac{E_{b}(V_{g})}{k_{B}T}\Big)
\end{equation}
\normalsize

where $A^*$ is the effective 3D Richardson constant ($A^{*}_{Si,3D} = 2.1 \times 120 \,A\cdot cm^{-2} \cdot K^{-2}$), $T$ is the temperature, $k_{B}$ is the Boltzmann constant and \textit{e} is the electronic charge. This will hold only when the cross-section size of the FinFET is large enough (i.e.: $W$, $H$ $>$ 20 nm) to be considered a 3D bulk system. In this study, $S$ is extracted for FinFETs with W(H) $\approx$ 25 nm (65 nm). When the $3D$ approximation is not true anymore (eg. $W$ or $H$ $\lesssim$ 20 nm), only $E_{b}$ and $\alpha$ can be correctly extrapolated \cite{Tet150}.

\section{Results and Discussion}\label{Sec:IV}

Two techniques to extract $D_{it}$ in n-Fin-FETs are presented.

\subsection{Method I}

\begin{figure}[t]
	\centering
		\includegraphics[width=3.5 in,height=1.8in]{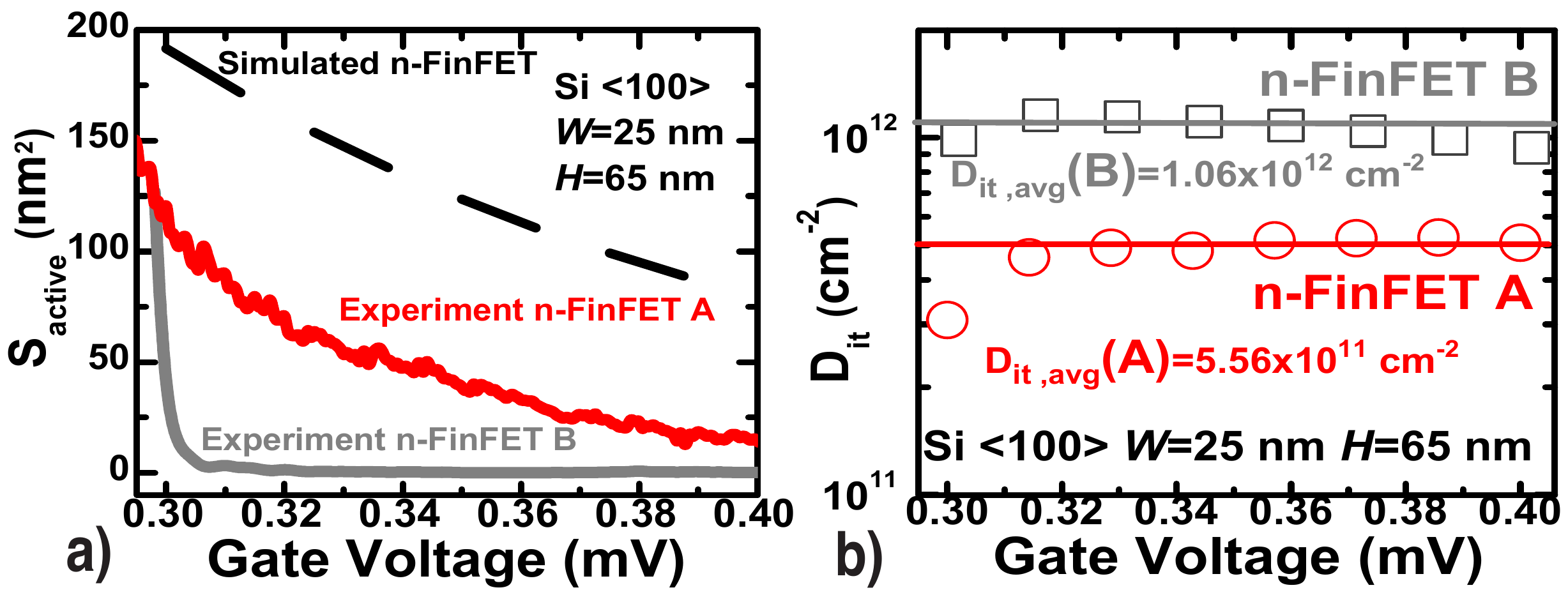}
	\caption{a) Experimental values of S for an n-FinFET device fabricated ($A$) using $H_{2}$ annealing step \cite{Lee186} and ($B$) without $H_{2}$ annealing. These curves are then compared with the simulated one. All the devices have $W$= 25 nm, $L$= 40 nm, $H$= 65 nm and an identical gate stack, but the use of different deposition systems has led to a shift of 0.15 V in the $V_T$ of ($B$), which is corrected in the figure for clarity. b) $D_{it}$ extracted for the two devices using Eq. \ref{trap_den1}. The lines show the average $D_{it}$ value ($D_{it,avg}$). Interface trap density is assumed to be constant for the top and the side walls of the FinFET which may not be always true \cite{Kap232}.}
	\label{fig:S_w5h65_match}
\end{figure}

The active cross-section in the undoped n-FinFETs is extracted from the temperature based conductance measurement using (\ref{Gbulk}) as outlined in Ref.~\cite{Tet150}.  Theoretically $S$ is extracted for two n-FinFETs with W/H = 25 nm/65 nm ($A$ and $B$) and the comparison of the simulated value of $S$ with experimental data is shown in Fig.~\ref{fig:S_w5h65_match}a. The simulations overestimate the value of $S$ due to the electrostatic screening of the channel from the gate due the interface trap charges ($\sigma_{it}$) present in these FinFETs \cite{Tet150,Kap232}. Based on the difference in the simulated $S_{sim}$ and the experimental $S_{expt}$, values, a method to extract $D_{it}$ in the FinFET devices is outlined (see table \ref{table2}). The method is based on the fact that the total charge in the channel at a given $V_{g}$ must be the same in the experiments and in the simulations. We therefore assume that the difference between the simulated ($\rho_{sim}$) and the trap charge density ($\rho_{it}$) is equal to the experimental charge density ($\rho_{expt}$) (see appendix). This leads to the following equation:

\begin{equation}
 \label{trap_den1}
\sigma_{it}(V_{g}) = \frac{\rho_{sim}(V_{g})S_{sim}(V_{g})}{e\cdot P}\left[\frac{\left[1-\frac{S_{expt}(V_{g})}{S_{sim}(V_{g})}\right]}{\left[1-\frac{S_{expt}(V_{g})}{W\cdot H}\right]}\right]\;[\#/cm^2]
\end{equation}
\normalsize

where  $P$ is the perimeter of the channel under the gate ($P = W+2H$). The extracted $D_{it}$ ($\approx$ $\sigma_{it}$) with $V_{g}$ (for W/H = 25 nm/65 nm), based on (\ref{trap_den1}), is shown in Fig.~\ref{fig:S_w5h65_match}b. The $D_{it}$ value is almost constant with $V_{g}$ showing that all the traps are filled (and therefore justifying the assumption $D_{it}$ $\approx$ $\sigma_{it}$). The average value of interface trap density $D_{it,avg}$ is obtained as $5.56\times10^{11} cm^{-2}$ for device $A$ and $1.06\times10^{12} cm^{-2}$ for device $B$ (Fig.~\ref{fig:S_w5h65_match}b). The values for $D_{it,avg}$  compare quite well with the experimental $D_{it}$ values for two different $L_{ch}$ devices from Ref.\cite{Kap232} (table \ref{table2}). The validity of these results are also supported by the fact that the obtained interface trap density values are different for devices with different surface treatment, as it is well known that the hydrogen anneal step during fabrication greatly improves device characteristics  and reduces the interface trap density \cite{Lee186}. As expected, for device $B$ we find much higher $D_{it,av}$ compare to device $A$.

\subsection{Method II}

By assuming that the surface potential ($\Psi_{s}$) and the $E_b$ respond equaly to $V_g$ \cite{Tet150, sze1}, a second extraction approach for $D_{it}$ is developed. Simple calculations (see \cite{sze1} and the appendix) result in an integrated trap charge density:

\begin{equation}
 \label{trap_den2}
\sigma_{it} = \frac{C_{ox}}{e}\int\frac{1}{\alpha_{sim}(V_{g})}\left[\frac{\alpha_{sim}(V_{g})}{\alpha_{expt}(V_{g})}-1\right]dV_{g}\;[\#/cm^2]
\end{equation}
\normalsize

with $C_{ox}$=0.0173 $F/m^2$ ($C_{ox}$ is assumed to be the same for all the devices since the oxide thickness is the same in all the devices). This method depends on the valid range of the $V_g$ used for integration in Eq. (\ref{trap_den2}). The limits are set from the $V_g$ point where $\alpha_{sim}$ $\approx$ 1 till the threshold voltage ($V_{T}$) of the FinFET (Fig.~\ref{fig:figure3}), after the flat-band shift ($\Delta V_{FB}$) of the simulated curve. All the extracted $D_{it}$ $\approx$ $\sigma_{it}$ from Method II are shown in table \ref{table2}. In the calculation few assumptions were made; (1) the extra charge contribution is assumed to come completely from the interface trap charges ($\sigma_{it}$) and any contribution from the bulk trap states have been neglected, (2) this method of extraction works best for undoped channels since any filling of impurity states is neglected in the calculations.

\section{Conclusion}\label{Sec:V}

\begin{figure}[t]
	\centering
		\includegraphics[width=3.55 in,height=1.6 in]{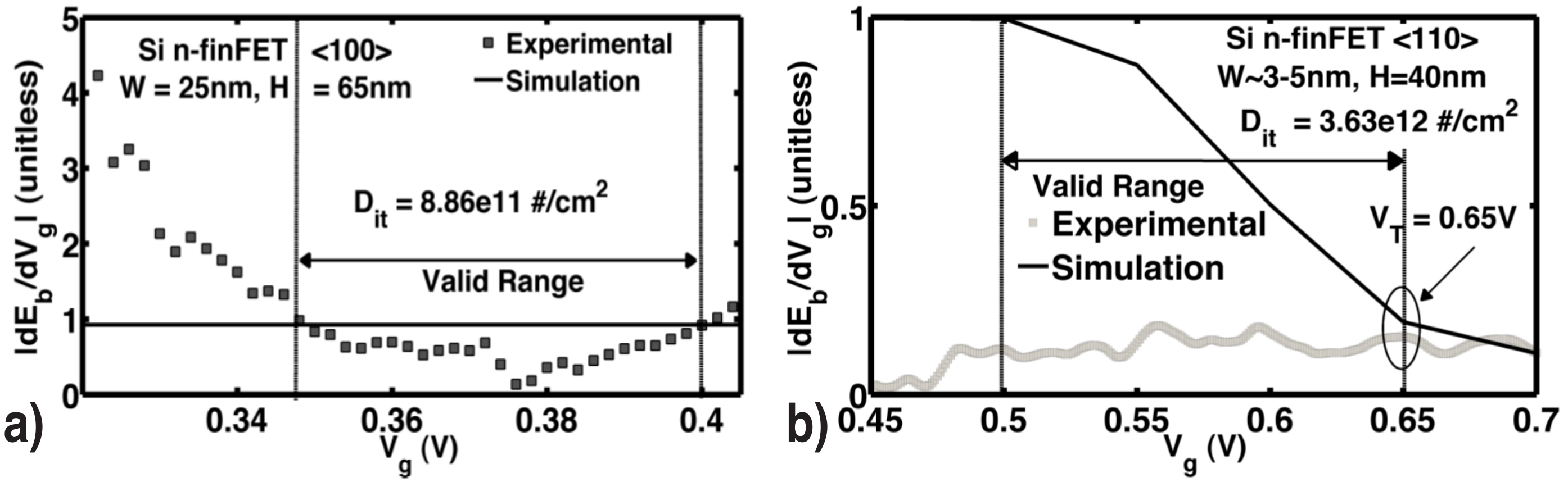}
	\caption{a) Simulated and experimental $\alpha=dE_{b}/dV_{g}$ for FinFET $B$ having $W$ = 25 nm, $L$ = 40 nm, $H$ = 65 nm, $<$100$>$ channel orientation and no hydrogen annealing step \cite{Lee186}. The observed mismatch is due to the presence of $D_{it}$, since excessive interface traps can screen the action of gate on $E_{b}$. Using Eq. \ref{trap_den2} it is then possible to calculate $D_{it}$ $\cong$ $\sigma_{it}$. It is important to notice that the numbers obtained using this method II for device $B$, compares well with the value obtained for the same device using the method I (as shown in table \ref{table2}). This gives a strong indication of the complementarities between the two approaches that have been introduced in this study. b) Simulated and experimental $\alpha=dE_{b}/dV_{g}$ for FinFET ($F$) having $W$ $\sim$ 3-5 nm, $L$ = 40 nm, $H$ = 40 nm, and $<$110$>$ channel orientation. The increase (X2) of $D_{it}$ compared to the devices with larger $W$ (i.e.: $D$ and $E$), may be attributed to the excessive etching required to make thinner fins.}
	\label{fig:figure3}
\end{figure}

A new $D_{it}$ determination methodology for state-of-the-art n-FinFETs is presented. Two complementary approaches provide (a) the gate bias ($V_{g}$) dependence of $D_{it}$ (Method I) and, (b) the total $D_{it}$ (Method II); both found consistent with each other. The following trends are observed; (i) devices fabricated without hydrogen annealing step, with smaller $W$'s and with $\langle$110$\rangle$ channel orientation show higher $D_{it}$ compared to the other devices, (ii) by comparison of the value of $D_{it}$ obtained for device $B$ in the two approaches (Fig.~\ref{fig:S_w5h65_match} and Fig.~\ref{fig:figure3}) and the value of $D_{it}$ obtained for two identical devices ($D$ and $E$) using the same approach (Method II), compatibility and reproducibility of the methods are demonstrated. The reported trends are similar to the one suggested in the literature \cite{Lee186}.

 \begin{table}[!t]

\begin{center}
\resizebox{8.8cm}{1,90cm}{
\begin{tabular}{|c|c|c|c|c|}
\hline
 & & &  &  \\
 \textbf{n-FinFET label} &\textbf{Approach}&\textbf{\textbf{$D_{it}$ ($10^{11}$} \textbf{$cm^-2$} )}& \textbf{FinFET type} &\textbf{Remarks}\\
 & & & &  \\
\hline

L=140nm \cite{Kap232}& Charge Pumping  &1.725 & Gated-diode FET \cite{Kap232} &$--$\\
\hline
L=240nm \cite{Kap232}& Charge Pumping  &2.072 & Gated-diode FET \cite{Kap232}  &$--$\\
 \hline  
 A &Method I $$&5.560&Standard FET&$H_2$ anneal, Low $D_{it}$\\
 \hline   
 B &Method I &10.6&Standard FET&\textbf{No $H_2$ anneal, higher $D_{it}$}\\
 \hline   
 B &Method II &8.86&Standard FET&\textbf{No $H_2$ anneal, higher $D_{it}$}\\
 \hline   
 C &Method II &9.26&Standard FET& Thin fin width, more etching, Higher $D_{it}$\\
 \hline    
 D (E) &Method II &18.31 (15.3)&Standard FET& $<$110$>$, higher bond density, Higher $D_{it}$\\
\hline  
F   &Method II &36.3&Standard FET&Same as C+D, Much Higher $D_{it}$\\
\hline  
\end{tabular}}
\end{center}
 \caption{$D_{it}$ obtained previously (\cite{Kap232}) compared with our results. It is possible to infer three expected trends. a) In good agreement with ref. \cite{Lee186}, the hydrogen annealing step substantially reduces $D_{it}$, b) The scaling of the $W$ of the devices (i.e.: from $A$ to $C$ or from $D$ ($E$) to $F$) increases the presence of interface traps and c) The change in the orientation of the channel (and therefore the sidewall surface where the interface traps are formed) from $<$100$>$ (device $A$ or $C$)  to $<$110$>$ (device $D$ ($E$) or $F$) remarkably increases the presence of traps. Unlike in \cite{Kap232}, the presented methods do not require any special device structure, which makes them easily applicable.}
\label{table2}
\end{table}

\section{Acknowledgements}\label{Sec:VI}
G. C. Tettamanzi and S. Rogge are with the Delft University of Technology, 2628 CJ Delft, The Netherlands. A. Paul, S. Lee, S. R. Mehrotra and G. Klimeck are with the Network for Computational Nanotechnology and the School of Electrical and Computer Engineering, Purdue University, West Lafayette, IN-47906, USA. N. Collaert and S. Biesemans are with IMEC, 3001 Leuven, Belgium. G.C.T. and S.R. acknowledge FOM and the European Community Seventh Framework under the Grant Agreement nr: 214989-AFSiD for the financial support. A.P., S.L., S. R. M. and G.K acknowledge the financial support from SRC, FCRP-MSD and NSF. Computational resources provided by nanoHUB.org is also acknowledged. G.C.T. acknowledge the kind hospitality extended by Prof. A. Di Carlo at the University of Tor Vergata, Rome, during the preparation of this manuscript. *G.C.T. and A.P. contributed equally to this work. (Emails:~ giuseppe.tettamanzi@gmail.com, abhijeet.rama@gmail.com).%

\newpage


\section{Appendix}\label{sec:0}
As outlined in Ref.~\cite{Tet150}, in undoped n-FinFETs the active channel cross-section area ($S$) and the source-to-channel barrier height ($E_{b}$) can be extracted from the temperature based conductance measurements. Furthermore, in these devices, the differences between the values of the experimental and the simulated $S$ and $\vert \partial E_{b}/\partial V_{g} \vert$ can be used for the extraction of interface traps density ($D_{it}$).

In this appendix, the details about the extraction procedure of $D_{it}$ are outlined. The material is organized in the following sections. Interface trap extraction using difference in $S$ is outlined in Sec. \ref{sec:I} along with the assumptions. Section \ref{sec:II} outlines the procedure for $D_{it}$ extraction using the differences between experimental and simulated gate-to-channel coupling values ($\alpha=\vert \partial E_{b}/\partial V_{g} \vert$) along with  the assumptions. Conclusions are given in Sec. \ref{Sec:V}.

\subsection{Method I} \label{sec:I}

Based on the difference between the simulated and the experimental active channel cross-section area values  ($S_{sim}$ and $S_{expt}$ respectively), a method to extract $D_{it}$ in FinFET devices is outlined. 

As the total charge in the channel at a given $V_{g}$ must be the same in the experiments and in the simulations (charge neutrality), the following equation is obtained:

\begin{equation}
\label{charge_equality}
S_{sim}\cdot L_{ch} \cdot \rho_{sim} = S_{expt}\cdot L_{ch} \cdot \rho_{expt} + e\cdot\sigma_{it}\cdot L_{ch} \cdot P
\end{equation} 
\normalsize

where  $L_{ch}$ is the channel length, P is the perimeter of the channel under the gate (P = $W+2H$ as show in Fig. 1 of the main paper), $W$ is the width of the channel, $H$ is the height of the channel, $\rho_{sim}$ ($\rho_{expt}$) is the simulated (experimental) charge density, e is the electronic charge and $\sigma_{it}$ the number of trap charges at the interface. Locally it can be assumed that $\rho_{expt}$ can be obtained from $\rho_{sim}$ and $\sigma_{it}$ using Eq. (\ref{rhoexp}):

\begin{equation}
 \label{rhoexp} 
 \rho_{expt} = \rho_{sim} - \rho_{it} = \rho_{sim} - (e\cdot\sigma_{it} \cdot P)/(W\cdot H)
\end{equation}
\normalsize

Using Eq. (\ref{charge_equality}) and (\ref{rhoexp}) a final expression for $\sigma_{it}$ is obtained as,

\begin{equation}
 \label{trap_den1}
\sigma_{it}(V_{g}) = \frac{\rho_{sim}(V_{g})\cdot S_{sim}(V_{g})}{e\cdot P}\left[\frac{\left[1-\frac{S_{expt}(V_{g})}{S_{sim}(V_{g})}\right]}{\left[1-\frac{S_{expt}(V_{g})}{W\cdot H}\right]}\right] [\#/cm^{2}]
\end{equation}
\normalsize

from this, the interface trap density can be extrapolated using the approximation $D_{it}$ $\approx$ $\sigma_{it}$.

\subsubsection{Assumptions in Method I} 

In the calculation of $\sigma_{it}$ ($D_{it}$) few assumptions were made. The extra charge contribution completely comes from the interface trap density ($D_{it}$) and any contribution from the bulk trap states have been neglected. Also all the interface traps are assumed to be completely filled which justifies the fact that we can assume that $D_{it}$ $\approx$ $\sigma_{it}$. This method of extraction works best for undoped channel since any filling of the impurity/dopant states is neglected in the calculation. Also the interface trap density is assumed to constant for top and side walls of the FinFET which is generally not the case \cite{Kap232}.

\subsection{Method II}\label{sec:II}

\begin{figure}[b!]
\includegraphics[width=3in,height=2.2in]{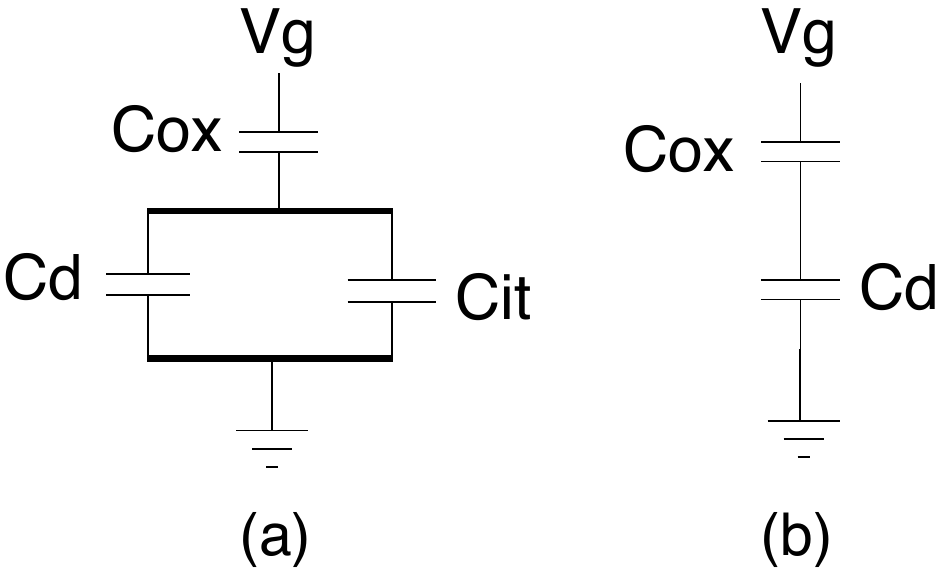}
\caption{Equivalent circuits (a) with interface-trap capacitance ($C_{it}$) and (b) without interface capacitance. $C_{d}$ and $C_{ox}$ are the depletion and the oxide capacitance, respectively. The idea for this equivalent circuit is obtained from page 381 in Ref. \cite{sze1}. }
\label{C_trap_fig}
\end{figure}

Based on the difference between simulated and experimental value of $\alpha$, a second method for the extraction of $D_{it}$ has been demonstrated. Starting from the equivalent capacitance model for a MOSFET with and and without the interface traps, as shown in Figure \ref{C_trap_fig}, the basic idea comes from Eq.(38) on page 383 in Ref. \cite{sze1} which gives,

\begin{equation}
\label{dEb_dEvg}
\vert \frac{\partial E_{b}}{\partial V_{g}} \vert = 1 - \frac{C_{tot}}{C_{ox}},
\end{equation} 

where $C_{tot}$ and $C_{ox}$ are the total and the oxide capacitance, respectively. For the two cases shown in Fig. \ref{C_trap_fig} the total capacitance is given by,

\begin{eqnarray}
\label{exp_c} C^{exp}_{tot} & = & \frac{C_{ox}\cdot (C_{d}+C_{it})}{C_{d}+C_{ox}+C_{it}}, \\
\label{sim_c} C^{sim}_{tot} & = & \frac{C_{d}\cdot C_{ox}}{C_{d}+C_{ox}},
\end{eqnarray} 

with $C_{d}$ being the depletion capacitance and  $C_{it}$ the interface capacitance. Equation (\ref{exp_c}) and (\ref{sim_c}) represent the total capacitances in the experimental and in the simulated device under ideal condition without any interface traps, respectively.

By combining Eq. (\ref{dEb_dEvg}), (\ref{exp_c}) and (\ref{sim_c}) and after some mathematical manipulations, we obtain,

\begin{equation}
\label{alpha_relation}
\frac{1}{\alpha_{exp}}  = \frac{1}{\alpha_{sim}} + \frac{C_{it}}{C_{ox}}, 
\end{equation}

where of course $\alpha_{exp/sim} = \vert \partial E^{sim/exp}_{b}/ \partial V_{g} \vert$.

By manipulating Eq.(\ref{alpha_relation}) it is possible to obtain Eq. (\ref{cit}):

\begin{eqnarray}
\label{cit}
C_{it} & =  & C_{ox} \cdot \Big(\frac{1}{\alpha_{sim}}\Big)\cdot\Big[\frac{\alpha_{sim}}{\alpha_{exp}}-1\Big], 
\end{eqnarray}

which can be associated with Eq. (\ref{citbis}) \cite{sze1};

\begin{eqnarray}
\label{citbis}
C_{it} & =&   e \cdot \frac{\partial \sigma_{it}}{\partial V_{g}}.
\end{eqnarray}

In Eq. (\ref{cit}) all the values are dependent on $V_g$ except $C_{ox}$. Integrating Eq. (\ref{citbis}) with regard to $V_{g}$ the final expression for the integrated interface charge density in the FinFETs is,

\begin{equation}
\label{method2_eq}
\sigma_{it} = \frac{C_{ox} }{e}\cdot \int_{V_1}^{V_2=V_{T}} \Big(\frac{1}{\alpha_{sim}(V_{g})}\Big)\cdot\Big[\frac{\alpha_{sim}(V_{g})}{\alpha_{exp}(V_{g})}-1\Big] dV_{g} [\#/cm^{2}],
\end{equation}

where $V_{T}$ is the threshold voltage of the FinFET and $V_1$ is the $V_{g}$ at which $\alpha_{exp} = 1$. This is the integration range for Eq. (\ref{method2_eq}) in the sub-threshold region.

\subsubsection{Assumptions in Method II}

While deriving the final equation for method II some assumptions were made. The first most important assumption is that 
the rate of change of the surface potential ($\Psi(V_{g})$) is same as $E_{b}$ with  $V_{g}$. The extra charge contribution completely comes from the interface trap density ($\sigma_{it}$) and any contribution from the bulk trap states have been neglected. Also all the interface traps are assumed to be completely filled which means $\sigma_{it}$ = $D_{it}$. This method works best when the change in DC and AC signal is low enough such that the interface traps can follow the change \cite{sze1}.

\subsection{Conclusion}\label{Sec:V}

The detailed calculation for both the trap extraction methods have been provided along with the assumtions made to obtain the final equations. The important point to note here is that, method I provides the variation in interface trap charges with the gate bias however, method II provides the total interface trap charge density within a range of gate bias. The application and the results obtained from these methods are provided in the main paper. 

\bibliographystyle{IEEEtran}


\end{document}